%% file: cern2011-001_nopdfx.tex
\documentclass[twoside,11pt]{cernrepm}
\newcommand{\BLKP}{% add blank page when on even page after PDF file
  \ifthenelse{\isodd{\value{page}}}{\relax}{\mbox{}\thispagestyle{empty}\newpage}}
\pdfoutput=1
\usepackage[T1]{fontenc}
\begin{document}
\pagestyle{empty}
\include{title}

\pagestyle{plain}
\pagenumbering{roman}
\setcounter{page}{3}
\include{frontmatter}

\cleardoublepage
\setcounter{page}{1}
\pagenumbering{arabic}
\include{adamsmono}

\end{document}

%% file: title.tex
\thispagestyle{empty}
\setlength{\unitlength}{1mm}
\begin{picture}(0.001,0.001)
%\graphpaper(0,-280)(210,290)
\put(120,13){CERN--2011--001}
\put(120,8){31 January 2011}
\put(0,-45){\includegraphics[width=15cm]{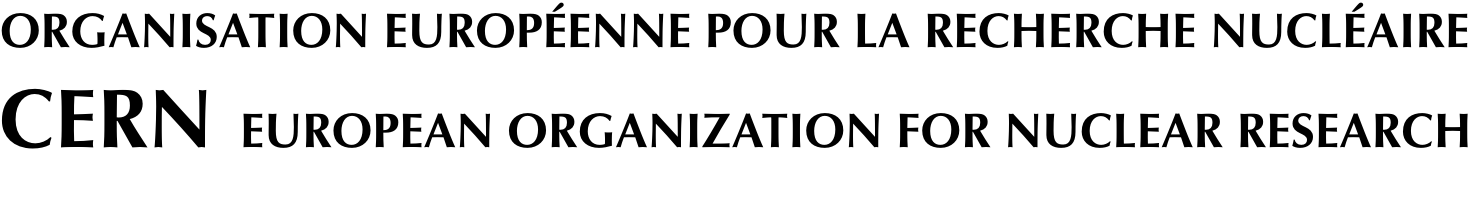}}
\put(68,-100){\makebox(0,0){\Large\bfseries Sir John Adams: %
         his legacy to the world of particle accelerators}}
\put(68,-115){\makebox(0,0){\Large John Adams Memorial Lecture 2009}}
\put(68,-135){\makebox(0,0){\Large E. J. N. Wilson}}
\put(68,-142){\makebox(0,0){\large John Adams Institute, University of Oxford, UK}}
\put(68,-240){\makebox(0,0){GENEVA}}
\put(68,-245){\makebox(0,0){2011}}
\end{picture}
\newpage
\thispagestyle{empty}
\mbox{}\\
\vfill
\begin{flushleft}%\large
\begin{tabular}{@{}l@{~}l}
  ISBN & 978--92--9083--356-7 \\% for CERN-2011-01
  ISSN & 0007--8328\\ % Non-CERN Physics School Publication
%    ISSN & 0531--4283 \\ % CERN School 
%    ISBN & 978--92--9083--343--7 % for CERN-2010-01
%    ISBN & 978--92--9083--353--6 % for CERN-2010-02
\end{tabular}\\[1mm]
Copyright \copyright{} CERN, 2011\\[1mm]
\raisebox{-1mm}{\includegraphics[height=12pt]{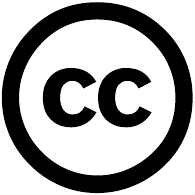}}
 Creative Commons Attribution 3.0\\[1mm]
Knowledge transfer is an integral part of CERN's mission.\\[1mm]
CERN publishes this report Open Access under the Creative Commons
Attribution 3.0 license (\texttt{http://creativecommons.org/licenses/by/3.0/})
in order to permit its wide dissemination and use.\\[3mm]
This monograph should be cited as:\\[1mm]
E. J. N. Wilson, Sir John Adams: his legacy to the world of particle accelerators, \\
John Adams Memorial Lecture, 2009, CERN-2011-001 (CERN, Geneva, 2011).\\[3mm]
\end{flushleft}

%% file: frontmatter.tex
\begin{center}
\mbox{}\\[3cm]
\bfseries\Large Abstract\\[1cm]
\end{center}

\noindent
John Adams acquired an unrivalled reputation for his leading part in
designing and constructing the Proton Synchrotron (PS) in CERN's early
days. In 1968, and after several years heading a fusion laboratory in
the UK, he came back to Geneva to pilot the Super Proton Synchrotron
(SPS) project to approval and then to direct its construction. By the
time of his early death in 1984 he had built the two flagship proton
accelerators at CERN and, during the second of his terms as
Director-General, he laid the groundwork for the proton--antiproton
collider which led to the discovery of the intermediate vector boson.
How did someone without any formal academic qualification achieve
this? What was the magic behind his leadership?  The speaker, who
worked many years alongside him, will discuss these questions and
speculate on how Sir John Adams might have viewed today's CERN.

\newpage
\BLKP
\cleardoublepage
\setcounter{tocdepth}{3}
\tableofcontents
\newpage
\BLKP
\cleardoublepage

%% file: adamsmono.tex
\section{Introduction}

Twenty-five years ago, in the year of John Adams's death, Edoardo
Amaldi gave the first talk in this series of John Adams Memorial
Lectures~\cite{ref1}.  Amaldi's subject was, like mine, the life of
this great man. In 1959, in the very early days of CERN, Amaldi had
recruited John Adams to build the Proton Synchrotron (PS) and had
remained his friend and supporter throughout his career. His account
was from the viewpoint of a senior figure in European accelerator
science.

My own account is written from the very different viewpoint of a
member of John Adams's team. My personal experience of the man dates
from 1969 when he returned to CERN for a second time as Project Leader
Designate of the 300~GeV Machine (or Super Proton Synchrotron (SPS) as
it was to become). I was a research fellow at CERN when he recruited
me as his technical assistant. I was given the job of adapting the
lattice of the SPS and coordinating its design to the point that, in
1971, CERN's Member States were finally able to approve the project
and agree that it should be built at CERN. I then continued to work in
day-to-day contact with John Adams as his Head of Parameters during
the design of the SPS and throughout its commissioning in 1976. I was
therefore fortunate enough to see him mastermind a huge project and
deal with the many obstacles that must be overcome in such an
endeavour. It is my hope that these two accounts complement each other
to give a full picture of the ingredients of his greatness.

John Adams was at the heart of CERN's proud boast that its
accelerators are finished on time and work reliably, and he should be
remembered as an example for all future machine builders and project
directors. In the course of writing this account, several questions
occurred to me. How did someone like John Adams without any formal
academic qualification achieve this? What was the style and method
behind his leadership? How did he achieve political success with the
Member States of CERN in turning the almost hopeless quest for
approval of the SPS to CERN's advantage? I will also attempt to
compare him with his US counterpart R. R. Wilson, and imagine what he
would now have to say about CERN's last 25 years. I believe the
answers to these questions will go a long way to understanding his
mastery of the field and I will therefore use italics to emphasize
them.

\section{How John Adams viewed building accelerators}

Let me return to the matter of John Adams's style
of building machines that were reliable and which cost no more than
promised. He attempted to summarize how he achieved this in some of his
final words to the CERN Council:

\emph{\ldots{}The question of how much flexibility to build into a
machine is obviously a matter of judgment, and sometimes the machine
designers are better judges than the physicists who are anxious to
start their research as soon as possible. But whatever compromise is
reached about flexibility, one should certainly avoid taking risks with
the reliability of the machine because then all its users suffer for as
long as it in service and the worst thing of all is to launch
accelerator project, irrespective of whether or not one knows how to
overcome the technical problems. That is the surest way of ending up
with an expensive machine of doubtful reliability, later than was
promised, and a physicist community which is thoroughly dissatisfied.
CERN, I am glad to say, has avoided this trap and has consistently
built machines which operate reliably, are capable of extensive
development, and have been constructed within the times promised and
within the estimated costs.}

\section{His first success}

\Fref[b]{fig1} is a picture of John Adams at a high point in his
career. It was taken on 25 November 1959, in the CERN
Auditorium---fifty years ago (almost to the day of this lecture) as he
announced to CERN Staff that the PS had accelerated beam to
24~GeV. The November 2009 issue of the CERN Courier~\cite{ref2}
contains an extract from a lively contemporary account by Hildred
Blewett of the previous night's excitement in the Control Room.

\begin{figure}[!ht]
\centering\includegraphics[width=194pt]{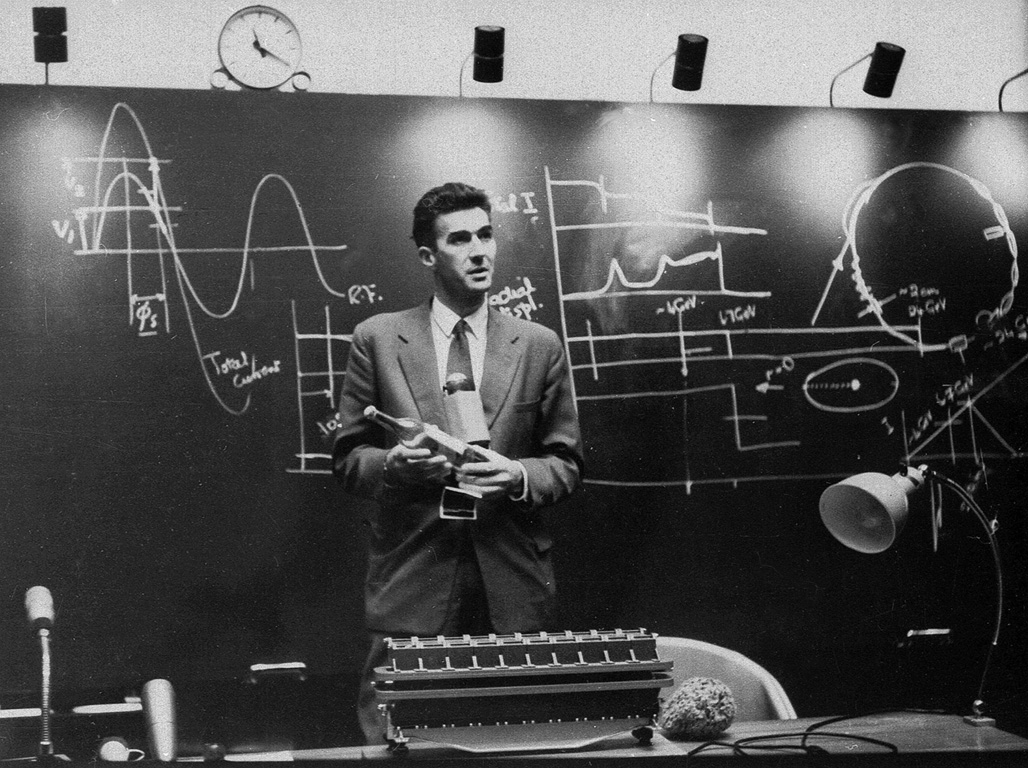}
\caption[]{John Adams announces that the PS had accelerated beam to 24~GeV}
\label{fig1}
\end{figure}

In his hand can be seen an (empty) vodka bottle, which he had received
from Yu. P. Nikitin with the message that it was to be drunk when CERN passed
Dubna's world record energy of 10~GeV. The bottle contains a Polaroid
photograph of the 24~GeV pulse ready to be sent to the Soviet Union.

\begin{figure}[!ht]
\centering\includegraphics[width=225pt]{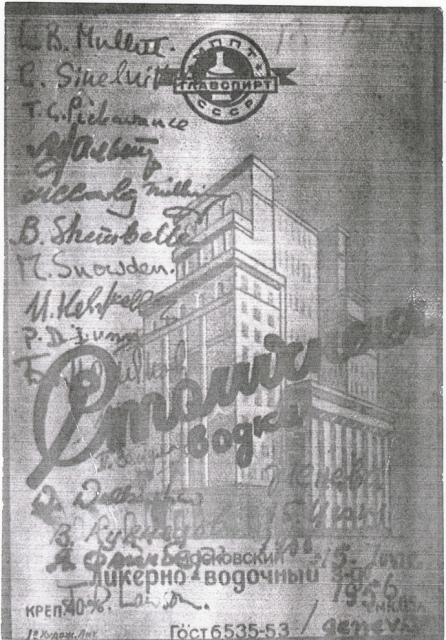}
\caption[]{The label of the vodka bottle (from the John Lawson
Archives~\cite{ref3})}
\label{fig2}
\end{figure}

The label which we see in \Fref{fig2} is itself a piece of
history\textemdash{}a testament to an international meeting at Dubna
some months earlier and one of the early cracks in the ice of the Cold
War. The names include Mullet, Pickavance, Crowley-Milling, Snowdon,
Lawson, and many others in Cyrillic script.

And this brings us to the first question: How did this young man of
33, without university education, come to lead such a project?
Certainly he was not coached in physics, mathematics or management at
a prestigious university. He had left school in 1936 without wishing
to go on to university. Rather, he sought practical employment as a
student apprentice at the Siemens Laboratories at Woolwich. He took a
Higher National Certificate (HNC) night school diploma in electronics
to become a member of the Institution of Electrical Engineers, but at
this point his formal education came to an end. When asked this
question many years later, John Adams said,
\emph{\textquotedblleft{}If university means that you learn from
capable men\textemdash{}I had ample opportunity}.\textquotedblright{}

\section{Telecommunications Research Establishment Malvern\textemdash{}his university}

He first began to meet these capable men when he joined up for the war
effort in 1940 and was posted to Telecommunications Research
Establishment (TRE) Swanage and later to Malvern where Radio Direction
Finding (RDF) or radar was under development.  The staff and advisors
of TRE included John Cockcroft, Robert Watson-Watt, Henry Tizard, Alan
Blumlein, Bernard Lovell, P. I. Dee, \mbox{W. E. Burcham,} and E.~D.~Fry. Many
of the accelerator builders of the post-war years were also there
including Hine, Crowley-Milling, Shersby-Harvie, Snowdon, Mullet,
Walkinshaw, and J.~D.~Lawson.

John Adams was in a group responsible for transmitter--receiver cells
and diodes for 3~cm radar. His boss and mentor then was Herbert
Skinner who had worked at the Cavendish Laboratory under
Rutherford. His contemporaries said Adams had an instinctive feeling
for what was needed\textemdash{}a comment that appears again and again
during his later career. Adams's roommate also said he was so good at
doing sketches he could design a complete three-dimensional circuit
layout on paper. It was during this time that he met Mervyn
Hine\textemdash{}later to be his closest collaborator in the design of
the PS, and Michael Crowley-Milling who was then working for
Metropolitan Vickers building linacs for medical purposes. Michael was
to become part of the Adams team that built the SPS and has written a
book about John Adams which I commend to you as a more complete
account than space allows me here~\cite{ref4}.

\section{Harwell}

\begin{figure}[!ht]
\centering\includegraphics[width=452pt]{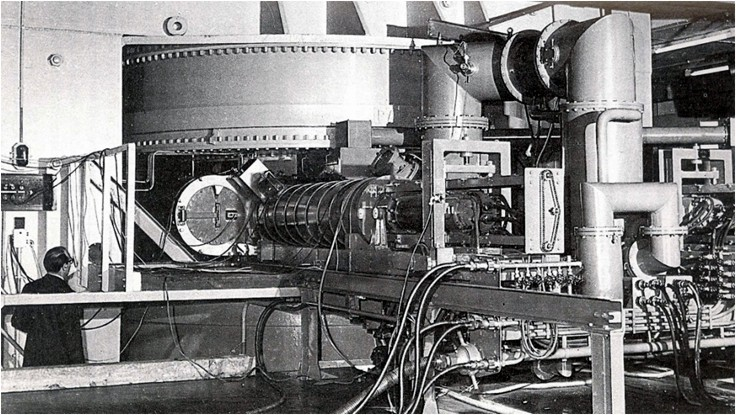}
\caption[]{The Harwell cyclotron}
\label{fig3}
\end{figure}

After the war, the Atomic Energy Research Establishment (AERE) Harwell
Laboratory was set up at the initiative of Sir John Cockcroft, Mark
Oliphant, and James Chadwick so that the contributions made by Britons
to the nuclear effort in the US might continue in Europe. As part of
this it was decided to build a 100-inch cyclotron
(\Fref{fig3}). Herbert Skinner, John Adams's boss from TRE, was in
charge of General Physics at Harwell and invited him to join the
project. He was to work under Gerry Pickavance who had been part of a
team that had already built a cyclotron at Liverpool University. At
the time, Gerry had a reputation for assuming an importance above his
station. It is said his Liverpudlian colleagues once nailed him to the
floor by the sleeves of his lab-coat to teach him a lesson in modesty.
As a Liverpool man myself, I can attest to this being quite within the
bounds of possibility, though by the time I met Gerry Pickavance as
Leader of the Rutherford Laboratory, he had obviously learned his
lesson in restraint, and had become an excellent senior manager who
was later to become a staunch supporter of John through the period
leading up to the SPS.

%WL: maybe error in inserting end mark: 
It was at Harwell that John cut his teeth on project management. The
Harwell cyclotron~\cite{ref5} was challenging\textemdash{}a
synchrocyclotron with 110-inch poles, closely modelled on Stan
Livingston's design for the Massachusetts Institute of
Technology. When Gerry spent months at a time visiting the US, John
was left in charge of everything except the RF systems. He found he
was taking more and more of the crucial design decisions himself as he
thoroughly worked his way through a multitude of sketches and
calculations as diverse as heat transfer and particle orbits. This was
a considerable responsibility for a young man and here perhaps is
another clue to his success as he \emph{seized this unusual
opportunity to develop his skills and experience.}

\begin{figure}[!ht]
\centering\includegraphics[width=120pt]{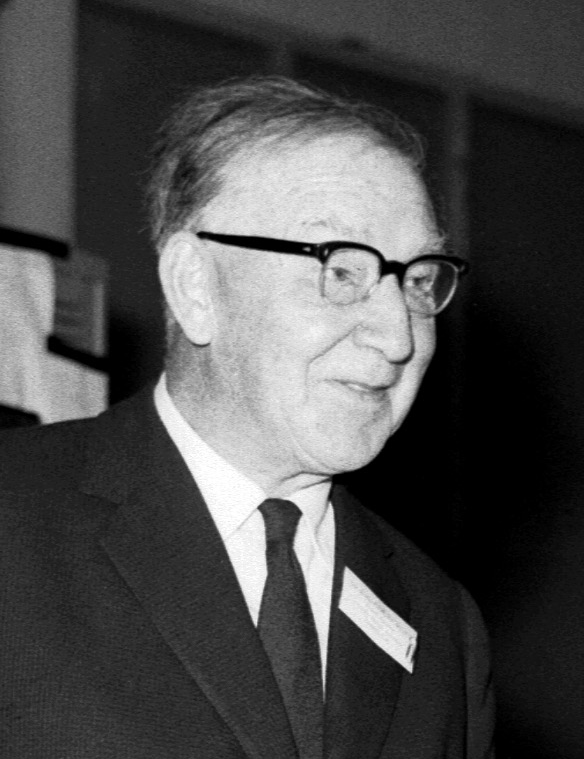}
\caption[]{Sir John Cockcroft}
\label{fig4}
\end{figure}

Even great men need a role model. For John Adams it was Harwell's
Director, Sir John Cockcroft, who had been awarded a Nobel Prize for
his atom-splitting at the Cavendish Laboratory in the 1930s
(\Fref{fig4}). Sir John Cockcroft was much revered by John who in
later life displayed a portrait of him behind his desk. Cockcroft is
said to have been a modest man who managed his team with quiet
authority. His management style was to let people get on with what
they were good at, but to show an almost daily interest in their
progress. He would often appear at the beginning of a day's work
behind the shoulder of a humble lathe operator to ask him
\textquotedblleft{}How is it going?\textquotedblright{} He gave his
staff considerable freedom to follow their own line, but would be
quick to support them by shouldering the responsibility, should they
need to be rescued. How different from the aggressively critical
attitude taught to today's managers who, all too often, are ready to
dismiss \textquoteleft{}the weakest link\textquoteright{} rather than
correct and reform. John's style was very much that of
Cockcroft\textemdash\emph{a style which I commend to those
who might wish to emulate him.}

\begin{figure}[!ht]
\centering\includegraphics[width=264pt]{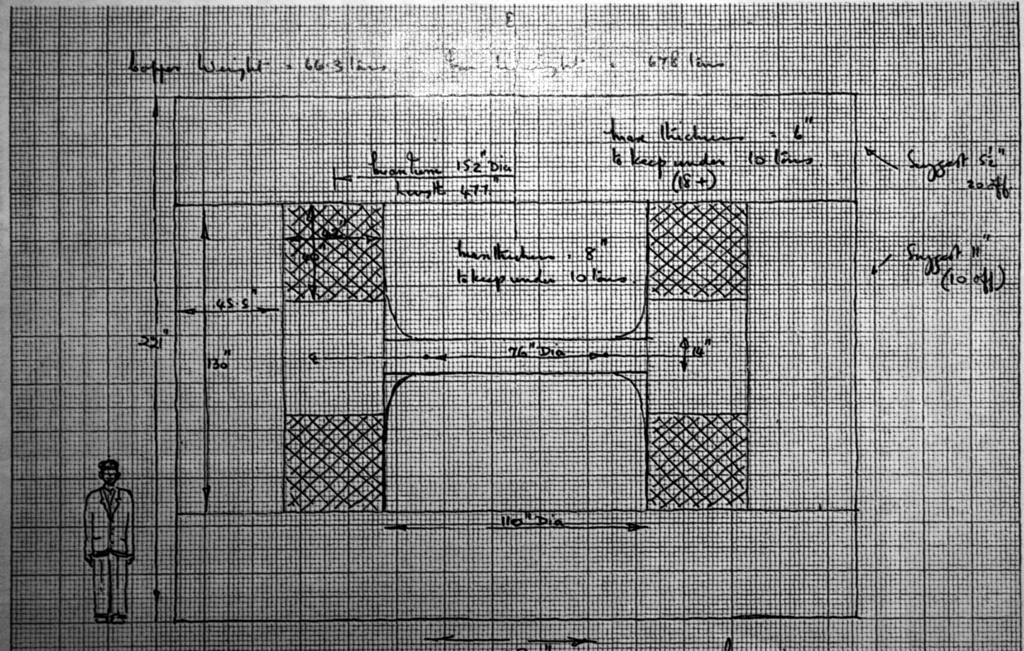}
\caption[]{A page from John Adams's meticulously kept Harwell notebook}
\label{fig5}
\end{figure}

Harwell was part of John's learning curve and it was there that he
first tasted failure when he tried to persuade Skinner to give him an
extra \textsterling{}50,000 to enlarge the yoke of the magnet and
reach a higher energy. (See \Fref{fig5} from his notebook.) He lost
the battle only to see the finished cyclotron end up with not quite
enough energy to produce the new `mesons' which it might have
discovered. This may have been in his mind as he later pressed for
400~GeV rather than 300~GeV for the SPS. Perhaps here he learned
another lesson\textemdash\emph{not to give up on something your gut
feeling tells you is correct.}

Once the Harwell cyclotron was finished, he had another
setback as he was reassigned to work on a fast breeder reactor. Knowing
very little nuclear physics, he had to work night and day to catch up
but found it frustratingly difficult. He became seriously depressed and
was sent away for six months by his wife Renie to stay with an uncle
who was a pig farmer. He returned in better spirits and with the
courage to discuss his future with Cockcroft. 

Cockcroft, who firmly believed in matching the man to the job, was
sympathetic and as a temporary measure set him to work on a klystron
together with Mervyn Hine. Soon after, Cockcroft saw a real chance to
rescue John by setting him off on an international venture that
brought him to CERN. Here he learned two more lessons---\emph{don't
force yourself to do things which do not match your skills }and
\emph{at crucial times seek help from your mentor.}

\section{CERN}

In May 1952 The CERN Council met for the first time in Paris. CERN's
initial idea for a Proton Synchrotron (PS) was a 10~GeV weak focusing
machine---a scaled-up version of the 3~GeV Cosmotron at Brookhaven in
the US, which had recently become the first proton synchrotron to
operate. A Norwegian, Odd Dahl, was the CERN PS Project leader
together with Frank Goward who was later to become his deputy in
Geneva. Very soon after this, in August 1952, Dahl, Goward, and Rold
Wider\"{o}e visited the Cosmotron and learned of the new idea of
strong focusing from Courant, Livingston, and Snyder. They returned to
immediately change the CERN plan for a 25~GeV alternating-gradient
machine.

At that time, the UK was suspicious of its continental
neighbours. After all, it had benefited from a vigorous partnership
with the US on nuclear matters during World War II and saw little
advantage in joining CERN. It fell upon Edoardo Amaldi and Cockcroft
to persuade a reluctant Ben Lockspeiser, then the UK minister in
charge of the Department of Scientific and Industrial Research, to
join. They also had to persuade Lord Cherwell, Churchill's scientific
advisor, to withdraw his objections to CERN. In this they eventually
succeeded.  Amaldi described in his first John Adams Memorial Lecture
how he then wished to meet some young British physicists and
engineers, whereupon Cockcroft brought John Adams to meet him at
lunch. Afterwards Amaldi had an extended interview with the young man
as they travelled by car to Harwell and chose to recommend John (and
Frank Goward) for places in the new team to build the PS. This set the
seal on the career that was to lead John to his first triumph. Amaldi,
in \Bref{ref1}, recalls that John was surprisingly ready to move to
Europe.  He already realised the role of international science in
keeping nations from warfare and wanted to be part of it. This seemed
one of John Adams's guiding principles destined to steer his life
towards CERN and later to world projects: `\emph{International common
ventures prevent wars.}' Frankly, as a child of wartime United
Kingdom, I appreciate how such thoughts were far in advance of their
time.

\begin{figure}[!ht]
\centering\includegraphics[width=415pt]{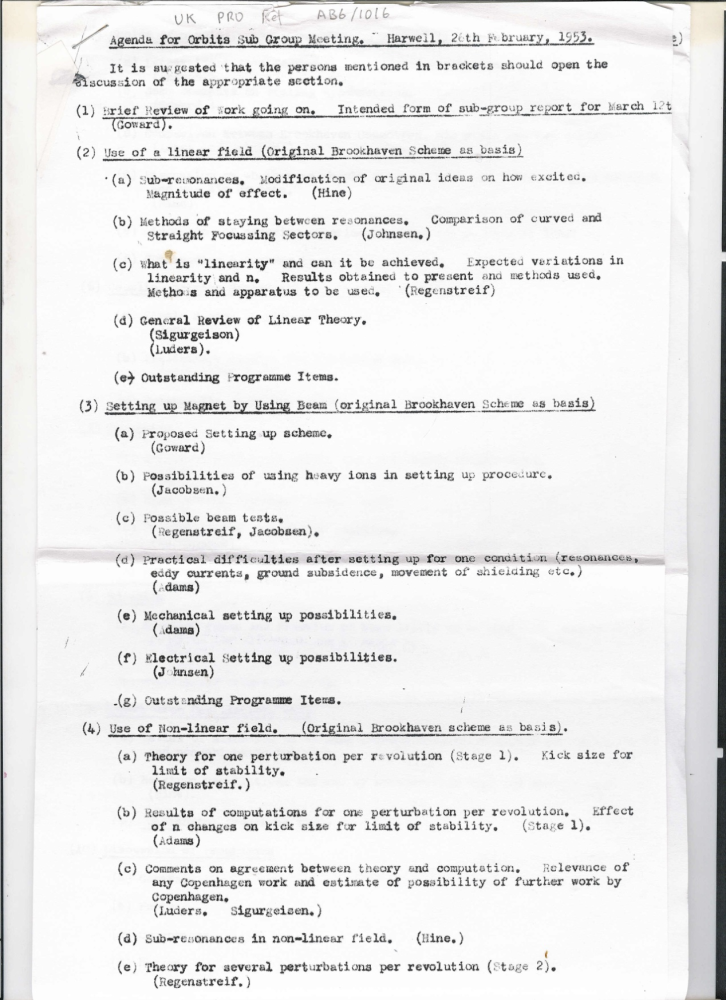}
\caption[]{Agenda of a meeting to decide PS parameters}
\label{fig6}
\end{figure}

The UK was therefore still not immediately a signatory
to CERN and, not for the last time, John found himself working on a
major European project without the support of his own government. But
Frank Goward and John Adams were seen as experts in circular machines
and they met frequently in Harwell and other laboratories to discuss the new
idea with the nascent PS team.

Among such discussions there was a crucial meeting at Harwell at the
end of 1952\textemdash{}just after Amaldi's visit, and before Adams
officially worked for CERN. Those at the meeting included
J.~D.~Lawson, Kjell Johnsen, Mervyn Hine and John Adams. It was not
minuted, but in \Fref{fig6} (from \Bref{ref3}) we see the agenda for a
subsequent meeting which gives a clue as to the contributions of the
various participants.

It was John's job to help resolve the many doubts there
still were about this decision to change to alternating-gradient
focusing. John Lawson had warned of the dangers of non-linear
resonances and Kjell Johnsen had to be persuaded that transition would
not be a problem. John and Mervyn Hine studied the non-linear resonances
driven by magnet imperfections using ACE, one of the first computers
available in the UK. It seemed that because of the high field gradient
(n-value) of the first design, magnet construction tolerances would
need to be unrealistically tight to avoid these resonances. Hine
writes: \textquotedblleft{}I remember at the end of the Harwell meeting
John summarized and took over. He stepped into the authority position
and wrote a summary on the blackboard in his wonderfully clear
left-hand writing.\textquotedblright{} In retrospect this seems to be a
crucial turning point at which \emph{he seized the opportunity to
assume authority over the new project's design}.

At subsequent meetings John was able to report that a set of
compromise parameters had been found. The n-value was to be reduced by
a factor 4 and the magnet aperture would have to be three times
larger\textemdash{}but still tiny compared with the Cosmotron. This
was typical of the kind of approach that John brought to the design of
accelerators. Each effect had to be analysed and calculated and its
effect on the chances of a successful outcome had to be balanced
against the need to be economical in construction. His notebooks
contained logical lists of arguments for and against each
compromise. He was to extend this careful elimination of all risk to
many other parts of the project and he recruited incomparable teams of
engineers to ensure the highest quality of design and construction.
Sometimes the workshops resembled a Swiss watch factory, but it paid
off and set the CERN standard for completing on time and within
budget. The secret was in the many long hours of discussion with
others and the analytical tool provided by his notebook.

\subsection{The PS Parameter Committee}

\begin{figure}[!ht]
\centering\includegraphics[width=328pt]{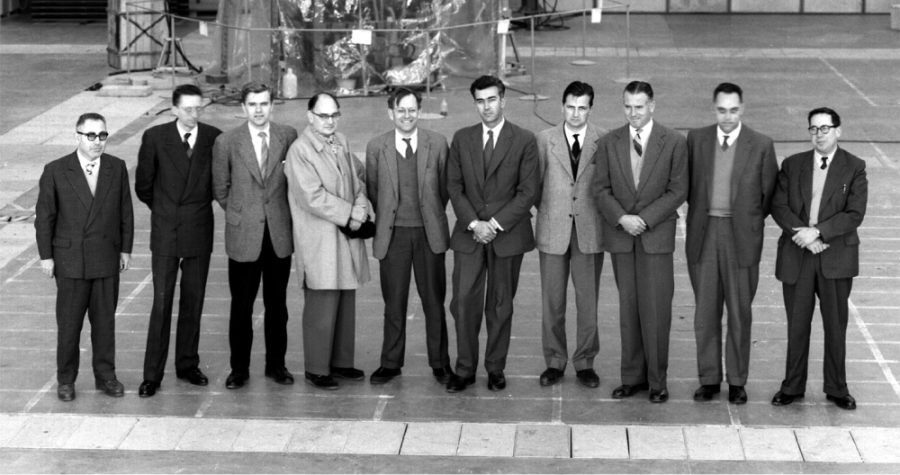}
\caption[]{PS Group Leaders--- From left to right we see
           Ed. Regenstreif, Pierre Germain, Kjell Johnsen, Arnold
           Schoch, Mervyn Hine, John Adams, Franco Bonaudi, Fritz
           Grutter, Kees Zilverschoon, and Colin Ramm}
\label{fig7}
\end{figure}

Not long after, in October 1953, the PS team gathered in premises lent
to them by the University of Geneva whilst awaiting the construction of the
first buildings on the new CERN site in Meyrin.  Goward was the
Project leader and he assembled a formidable team of experts to design
and construct the machine. In the photograph (\Fref{fig7}) we see
almost all the PS Group Leaders, each responsible for an aspect of the
machine. 

John Adams and Mervyn Hine were known at this time as the
\textquoteleft{}terrible twins\textquoteright{} using their
experience with earlier projects to enliven the proceedings of the
Parameter Committee which met once a week to put flesh on the bones of
the design. (Later I will say more about the Parameter Committee and
its role in Adams's method of project management.)

To Giorgio Brianti, arriving in November 1953, it was clearly
John who, chairing the Parameter Committee, masterminded the design and
construction. Brianti recalls that then John was already
\textquoteleft{}the Boss\textquoteright{}. Goward had fallen ill
and died in early 1954 leaving John to take over full responsibility
for the project, eventually steering to the moment of his triumph in
1959.

\section{Plasma research\textemdash{}the move to Culham}

Near the end of his first period at CERN he began to take
an interest in plasma research, attending a number of meetings with a
view to setting up another international organization. He was
interested in plasma accelerators and some experimental work started at
CERN. 

In 1958 a lot of plasma work in the UK was declassified and shared
with the Russians during a momentary thaw in the Cold War. It was in
the days of ZETA, the Harwell machine which prematurely announced the
dawn of energy from thermonuclear fusion. These hopes proved false,
but in spite of this, the UK was keen to set up a new laboratory at
Culham to develop the field. Even before the PS was finished, they
tried to persuade John to return as Director of this new
laboratory. He was anxious that his children attend schools in the UK
academic system to prepare them for university and he agreed to head
the new laboratory once the PS was finished. However, following the
death of Jan Bakker, then CERN's Director-General, in a plane accident
in April 1960, John Adams was appointed Director-General of CERN. His
return to the UK had to be delayed until he had not only finished the
work of getting the PS running properly but had seen the physics
programme take off.

\begin{figure}[!ht]
\centering\includegraphics[width=128pt]{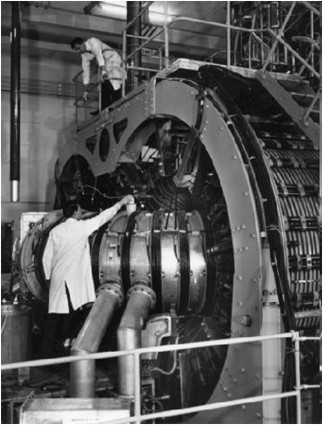}
\caption[]{ZETA}
\label{fig8}
\end{figure}

\begin{figure}[!ht]
\centering\includegraphics[width=128pt]{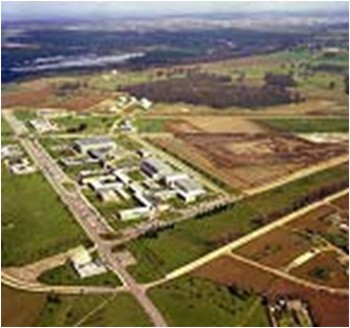}
\caption[]{The Culham Laboratory}
\label{fig9}
\end{figure}

In \Fref{fig8} we see ZETA, and in \Fref{fig9} the Culham laboratory
near Abingdon. He was to spend the next five years in the UK,
eventually being asked to advise Frank Cousins, a minister in the
government of Harold Wilson. I remember him later being very critical
of the quality of the administration over which Frank Cousins and
Anthony Wedgwood-Ben presided. John's advice was often not taken and
influenced government thinking only many years later. Frustrating as
this experience was, it left John with a clear idea of how politicians
worked and how they might or might not be influenced\textemdash{}an
experience that was to be invaluable in persuading Member States to
support moving the SPS to CERN.

\section{The 300~GeV machine and the ISR}

\begin{figure}[!ht]
\centering\includegraphics[width=172pt]{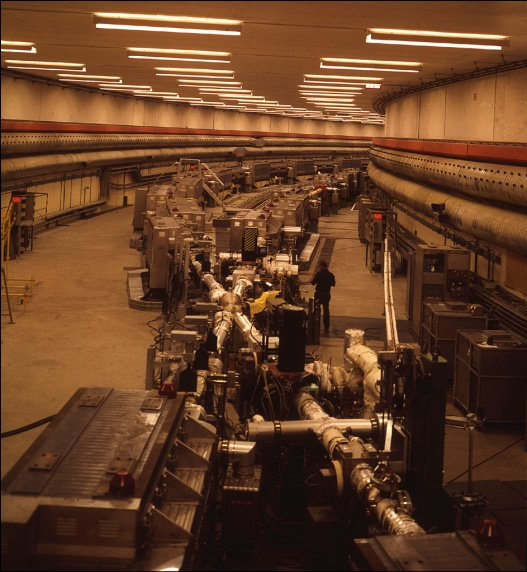}
\caption[]{The ISR}
\label{fig10}
\end{figure}

In 1960 when he was still Director-General and just before he left for
Culham, John recommended to Council, \textquotedblleft{}that CERN
should plan to build a machine to replace the Proton Synchrotron.  It
should be ready in 1970 therefore plans should be prepared for
consideration in 1962 or 1963.\textquotedblright{} A study group was
set up under Kjell Johnsen to look into the feasibility of a collider
(Interesecting Storage Rings, the ISR) and a 150--300~GeV proton
synchrotron. Council approved the ISR, appointing Johnsen to head its
construction (see \Fref{fig10}). At the same time a detailed design
study of a new proton synchrotron was published in a substantial
report \textquoteleft{}A Design Study for a 300~GeV Proton
Synchrotron\textquoteright{}~\cite{ref6}, commonly referred to as the
\textquoteleft{}Grey Book\textquoteright{}.

This machine proved to be a scaled-up version of the PS and
ISR, incorporating the lessons learned from their construction and
applied with all the conservatism that experienced engineers tend to
bring to new projects. It would have taken eight years to construct and
cost about 1800 MCHF.  

\subsection{Redesigning the 300~GeV machine}

It had not been the aim of Kjell Johnsen's team, who had written the
Grey Book, to be economical in either time or money. In the USA a
similar proposal was made for the 200 BeV accelerator, authored by
many who had contributed to CERN's Grey Book and incorporating many of
the same conventional features. After its publication, the
construction of the American machine was approved at the US Fermi
National Laboratory (Fermilab) near Chicago and work started under the
leadership of R. R. Wilson. He tore up the 200 BeV design and proposed
a much leaner design which could be constructed in only four years and
which would operate at 400~GeV\textemdash{}twice the energy originally
proposed. The most striking innovation was to change the lattice from
combined-function magnets to one in which the functions of bending and
focusing were performed by separate and quite different magnets.

The change from combined-function to separated-function lattice was
later to be so fundamental to securing approval for the SPS that it is
worth a short explanation. We recall that focusing in synchrotrons is
achieved by a field gradient across the mid plane of the
magnets. This, together with the centrifugal force on the particles,
forces errant particles which swing away either upwards, downwards, or
on either side of the ideal central orbit around the machine to be
pushed back towards the central orbit. In early synchrotrons and
cyclotrons this gradient was uniform around the machine. In the AGS,
PS, and ISR the sign of the gradient alternated from magnet to magnet
to produce a much stronger focusing effect called alternating-gradient
focusing. The magnets had tapered gaps between the poles so that they
both focused and bent the particles at the same time. The direction of
the taper alternated from magnet to magnet. Although the gradient
alternated in these machines, one kind of magnet combined the function
of bending and focusing which had a certain simplicity. These were the
magnets John and Kjell knew and loved from the PS and ISR.

However, in such magnets the central field, which determines the
central orbit and the radius of the machine, cannot be as high as it
is at the edge of the poles where saturation limits the field to 1.8
T. Allowing for the gradient, the field on the centre line of a
combined-function magnet can only be about 1.3 T. In the
separated-function idea there are two kinds of magnet:
\textquoteleft{}pure dipole\textquoteright{} bending magnets with
uniform field of about 1.8 T and special \textquoteleft{}pure
gradient\textquoteright{} quadrupole magnets to provide the focusing.
Bob Wilson had shown that, by using such a separated-function design,
there could be a considerable saving in total bending magnet
length\textemdash{}more than enough room to place special dedicated
quadrupole magnets to provide the focusing.\footnote
    {The first proposal of separate-function magnets (i.e., the
     separation of dipoles and quadrupoles) for an accelerator lattice
     was made by T. Kitagaki in 1953~\cite{RefKitagaki1953}. Among other
     implications, this separation allowed for smaller magnets and for
     the introduction of long straight sections without dipole
     fields.}

This change from combined to separated function happened in1967 just
when I had come to CERN on a fellowship and was encouraged by Roy
Billinge (then about to leave to build the Booster at Fermilab) to
join the Accelerator Research Department (AR) and work on improving
the {Grey Book}. Roy and I (we were both just 30) were both
\textquoteleft{}rebels with a cause\textquoteright{} convinced that,
by adopting some of the radical simplifications that Bob Wilson was
adopting for the Fermilab machine, CERN's 300~GeV machine would become
faster to build and cheaper. Roy went off to Fermilab while I taught
myself the rudimentary skills of lattice design and set about
designing a separated-function version of the Grey Book.

When I applied this to the 300~GeV machine, the energy jumped from
300~GeV to 400~GeV but when I showed this proudly to Kjell Johnsen and
Kees Zilverschoon (caretakers of the 300~GeV project, but still busy
finishing the ISR), I was told not to rock the boat. To be fair, the
main concern at that time was to decide it was to be built. Council
delegates spent much of the time viewing and reviewing the 22 and the
finally 5 possible sites scattered across Europe in countries who all
hoped to benefit from the local trade and prestige. It took the
arrival of John Adams to give the revised design the attention it
deserved.

In 1969 John Adams returned to CERN, appointed by the CERN Council to
lead the 300~GeV/SPS Project. Finding that I was the only full-time
person working on the lattice for the machine, he invited me to
several one-on-one discussions about the design of the new machine and
listened with enthusiasm to my separated-function version of the Grey
Book. As a very junior visiting fellow to CERN at the time, I was both
surprised and flattered by the attention of one of the
\textquoteleft{}great men of science\textquoteright{}. I expected it
would be too revolutionary and might seem to him to prejudice the
operational reliability of the machine, but it clearly fitted in with
what turned out later to be his grand scheme for CERN. We worked hard
on redesigning the machine, and our proposal is to be found in
\Bref{ref7}.

\subsection{Difficulties with the 300~GeV Project}

Just before John arrived in CERN for the second time, the UK had dealt
a blow below the belt to the
\textquoteleft{}300\textquoteright{}. The Labour government, who
were strapped for cash, were not very interested in pure science and
saw no financial advantage in the project even if their site were to be
chosen. In June 1968, Sir Brian Flowers announced to the CERN Council
that they should not count the UK among the participating countries.
This was just after John had given up his influential responsibilities
in London to move to CERN and become Project Director Designate. He found
himself once again playing a crucial role in starting a project without
the support of his own country.

Apart from the withdrawal of the UK from the new project
and the difficulties in settling on a site for the
\textquoteleft{}300 Machine\textquoteright{}, the new Project
Director Designate had to pay more than lip service to Fermilab ideas. Bob
Wilson was by then boasting a five-year construction time for his machine,
an almost unbelievable cost profile, and 400~GeV to boot. Adams was
under considerable pressure from certain German physicists and Citron
(of the PS days) to move away from the \textquotedblleft{}lavish
practices\textquotedblright{} of the PS and the ISR and take
heed of the wind of change blowing strong from Fermilab in Illinois. 

He had to cut the cost of the 300~GeV proposal without
sacrificing reliability, resolve the question of where it would be
built, and defuse the feeling that Member State opinions were not being
taken into account. I have no doubt that these aims were listed on the
first page of his notebook soon after he returned and it is clear to me
in retrospect that he lost no time in devising a strategy to find a way
through the minefield.  It can hardly have taken him too long, because
there seemed to be no preliminary exploration of blind alleys on the
way---and all of this from a man who \emph{seemed not to have made
up his mind about anything until he had heard all sides of the
argument}\textemdash{}masterly leadership by any standards. After the
event John Adams explained the difficulties he faced thus:

\begin{quotation}
``Looking back, I think one can discern a number of reasons why our
Member States hesitated to reach a decision on the 300~GeV Programme
in the form it was presented at that time.

In the first place the economic situation in 1969 for science in
general and nuclear physics in particular was very different from the
ebullient years around 1964 and 1965 when the 300~GeV Programme was
first put forward. It was evident that several Member States of CERN
and possibly all of them found the cost of the Programme too high
compared with their other investments in science and with the growth
rates in their total science investments, which had dropped from
figures around 15\% per annum in 1965 to a few per cent per annum
in 1969.

In the second place, the idea of constructing a second European
laboratory for nuclear physics remote from the existing one, which had
seemed attractive in 1965, looked inappropriate in 1969, particularly
since it implied running down the existing CERN laboratory when the new
one got under way.

In the third place, so many delays had occurred in the 300~GeV
Programme and the American machine was coming along so fast that an
eight-year Programme to reach experimental exploitation seemed too
long.

Fourthly, it turned out that choosing one site amongst five
technically possible sites presented non-trivial political problems for
the Member States of CERN.''
\end{quotation}

This quotation is typical of the point-by-point analysis
that he used to summarize his view on any argument. When he sat down in
the afternoon to update his notebook with his resolution of the
arguments presented to him by others, he would light his pipe and often
compose just such a summary. He found this kind of logical analysis led
him to the most reasonable and sensible decisions and, when presented
to others, was persuasive and almost irresistible in its clarity of
thought and logic. 

This is frankly not the kind of rhetoric that a politician
might use to sway a crowd but it is calm, reflective, designed to raise
the minimum of eyebrows, and, above all, be persuasive in its relentless
logic. Brian Southworth, then Editor of the CERN Courier once said:
\textquotedblleft{}John has the astonishing gift of delivering absolute
truth in the manner of Farmer Giles leaning over his gate to comment on
the weather.\textquotedblright{} 

The ideas seem just to have occurred to him after the
project was approved, as a clever way of summarizing, but he surely
arrived at these conclusions almost immediately he arrived at CERN as
Project Director Designate since it so well summarizes what everyone was
experiencing at that time\textemdash{}problems which only he knew how
to resolve. I believe he decided then to adopt it as a to-do list and,
playing his cards close to his chest, tackled each item in turn.

\begin{figure}[!ht]
\centering\includegraphics[width=133pt]{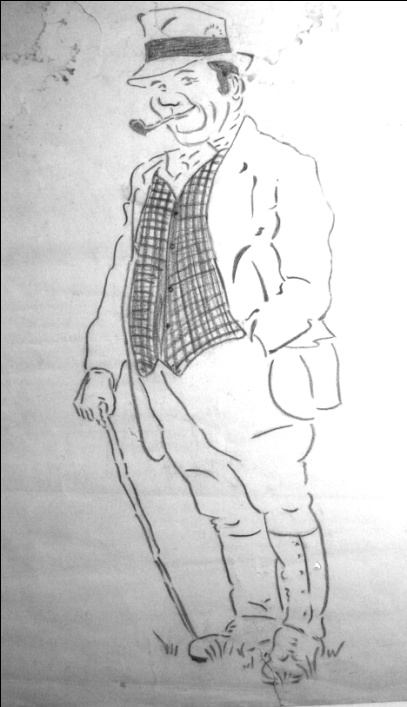}
\caption[]{\textquotedblleft{}Farmer Giles\textquotedblright{} (as sketched by John Adams)}
\label{fig11}
\end{figure}

\subsection{Designing the new machine}

His first step towards securing the CERN Council's approval
for the SPS was to set up a Machine Committee to involve as many of the
senior accelerator experts from Member States as he could. The choice
of this committee was masterly\textemdash{}that of a benign
Machiavelli. He needed the help of a number of his old PS group leaders
to ensure that high standards of engineering were not sacrificed for
the most important components, and to ensure the maximum probability of
success. Magnet, radio frequency, survey, and extraction were looked
after in this way. There were other major systems, among them the power
supply and the control system where he found those in Member States who
felt they could make an innovative contribution. Crowley-Milling's
control system based on mini computers from Norway was doubly salutary,
as were John Fox's power supplies based not on rotating machinery but
saturable reactors. Others were recruited to the committee to reassure
the sceptics in Germany and the UK that cost and manpower estimation
was done in the way they would like to see it. They were encouraged to
hang the redesign of the machine on this new separated-function lattice
that I had designed.  

John had headed the Parameter Committee in the days of the
PS and he clearly expected the Machine Committee to operate in the same
fashion to ensure the consistency of the design and the success of the
project. Each of the meetings followed an agenda which was principally
a series of reports by those responsible for the major systems of the
machine. Each system was worked out in greater engineering detail
following suggestions at the previous meeting. At the heart of
the business was keeping a list of all the relevant design data from
top energy, through number of bending and focusing magnets, their
length and peak field, the injection and extraction systems, together
with the frequency and voltage applied to the accelerating cavities, and
even the diameter of the tunnel and the load on the cooling system.
Each time anything was changed, its impact on cost, performance,
reliability, and of course implications for other systems would be
discussed with all the hardware specialists present. Any changes had to
be incorporated in a master list of parameters and in the lattice
design. For the design and later construction of the SPS, I was lucky
enough to be in charge of both parameters and lattice.  Of course it
was John, always at the head of the table, who presided. I sat at the
other end keeping the minutes. He encouraged me to ask questions which
would provoke discussion and reveal weaknesses in the design which
needed to be debated and resolved. This method of managing a project
had the great advantage that there was only one meeting at which
technical matters of accelerator design and engineering would be
decided in the presence of all component group leaders who might be
affected.

I remember the lively, heated discussions between new members and some
of the members from his past PS team who had moved on meanwhile to
build the ISR.  They were at pains to squash the ideas of Bob Wilson
in the United States, often expressed by this younger
\textquotedblleft{}upstart Wilson\textquotedblright{} (perhaps they at
first believed I was a relative) in their midst. Meetings were not
without their explosive exchanges\textemdash{}not surprising
considering my own brash inexperience. They also hoped to be asked to
build these components and did not want to make it hard for
themselves.

But having a variety of opinions put forward around the table suited
John's style of facilitating a meeting. He was able to judiciously
move the project from the Grey Book towards a leaner design without
apparently taking sides. John's moderate and reasonable interventions
were usually in the form of a simple
question. \textquotedblleft{}Wouldn't that mean
that\dots\textquotedblright{}, or as he turned to someone not already
part of the combat, \textquotedblleft{}How would that affect the
magnet/power supply/schedule?{\textemdash}How would these new magnets
look?\textemdash{}What tolerances would be necessary in their
construction?{\textemdash}Would they reduce cost?{\textemdash}How
would they compare with the PS and ISR? and How might one inject and
extract the beam?\textquotedblright{} In this way he would orchestrate
the discussion by asking for opinions until he heard one which matched
his new way of thinking. Then he would summarize the
\textquoteleft{}consensus\textquoteright{} he had sculpted for us and
define what was to be studied next. Of course there were, meanwhile,
many pipes of tobacco to be prepared after meticulous cleaning of the
instrument to provide a pretext for reflexion.

When all of this had been debated, I would be expected to ask
myself in the minutes of each meeting to accommodate new aspects of the
design in an ever increasing series of lattices, each with new sets of
parameters. 

Later, when we came to construct the SPS, the Machine Committee became
the Design Committee and the debates about magnets, cavities,
injection, extraction, power supplies, and civil engineering were
again heated. The more controversial decisions were often concerned
with the lattice (myself) and the magnets (represented by Roy
Billinge, recently returned from the US). Both of us had a preference
for the new ways of building synchrotrons pioneered at Fermilab which
were often at odds with the ideas of the more experienced members of
the team.  At the time the discussions in the SPS Design Committee
were taking place, the news from Fermilab was not good and it became
clear that their magnets had not been made to the standards of
electrical integrity established at CERN. But John was not to be put
off from taking what was best from Fermilab and imposing CERN
standards on its construction.

In managing the construction, John Adams followed closely
the precepts of his mentor Sir John Cockcroft. He gave his group
leaders, the members of his Monday Morning Design Committee,
considerable latitude to manage their own groups. His interest was
always on achieving performance goals on time and without over
expenditure. Subsequently I have heard it said that his budget was
generous compared with later machines. All I can say is that he made
strenuous efforts to build the SPS for much less than the unit costs
achieved in the PS and ISR days. True there was, wisely, a contingency
in the funding, but this was not needed for the SPS and at the end of
the construction was reallocated to provide a new North Experimental
Hall.

\emph{To have such a weekly meeting with the heads of
your hardware groups and have them inform everyone on progress in all
aspects of the machine seems so fundamental to John Adams's style that
would-be project leaders should depart from this practice at their
peril.}

\begin{figure}[!ht]
\centering\includegraphics[width=224pt]{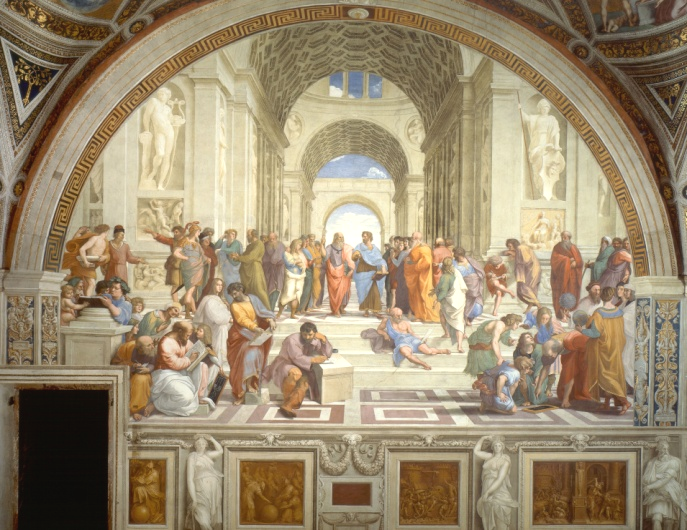}
\caption[]{Raphael's fresco `The School of Athens'}
\label{fig12}
\end{figure}

As I prepared this talk, I struggled to describe the
particular method that John Adams used to run a meeting. He hardly said
anything, but would steer the opinion of the members in the direction
he wanted simply by asking questions.  An Oxford philosopher friend
tells me this is exactly the method used by Socrates and Plato in  the
School of Athens (See \Fref{fig12}). \emph{Maieutics} (its name in Greek
means helping give birth\textemdash{}in this case to ideas) is a
disciplined questioning that can be used to pursue thought in many
directions and for many purposes, including: to explore complex ideas,
to get to the truth of things, to open up issues and problems, to
uncover assumptions, to analyse concepts, to distinguish what we know
from what we don't know, and to follow out logical implications of
thought. I suppose our budding project manager should read a bit of
Plato now and again\textemdash{}though I have no evidence that John
Adams did\textemdash{}he was probably hard-wired to act in this way.

\subsection{Design improvements}

But our narrative has run on and we must now return to the days when
his plan to secure the SPS for CERN was taking shape. He expected the
Machine Committee to think of improvements and to incorporate the new
ideas of Fermilab.

\begin{figure}[!ht]
\centering\includegraphics[width=248pt]{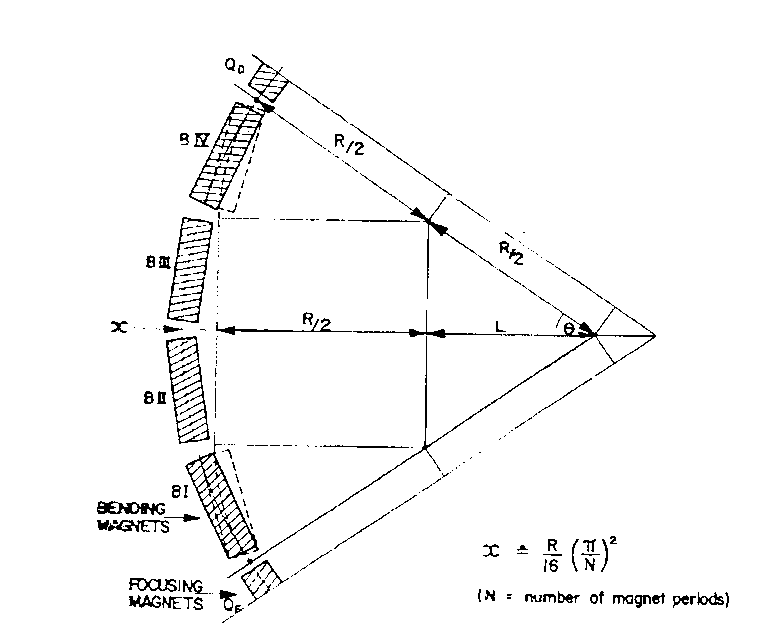}
\caption[]{One cell of a separated-function lattice (showing the missing magnet option)}
\label{fig13}
\end{figure}

The first system to be scrutinized was the lattice\textemdash{}the
pattern of magnets around the ring. Whether this is combined- or
separated-function, it always has to be consistent with the parameters
of the hardware. If it is decided to add more RF cavities to
accelerate faster or to increase beam capture efficiency, the lattice
has to be adjusted to make room for it.  The lattice determines the
dynamics of particles within the beam pipe, if it has many cells the
focusing will be stronger, the beam envelope smaller, and both magnet
dimensions and even that of the tunnel can be made smaller. Of course,
even if the logic of the mathematics tells you that the tunnel need
only be 150 cm in diameter, you can be sure that someone in the
committee will remind you that no one would be able to walk there, let
alone drive a lorry full of rock through it. In \Fref{fig13} we see
one cell of the lattice (out of 100 or so around the circumference).

\Fref[b]{fig13} also shows another new idea: missing magnets. If only
half of the bending magnets are built and installed in the first stage
but more money becomes available, you add the second half to double
the energy (from 200~GeV to 400~GeV). I'm not sure where this idea
came from\textemdash{}it was perhaps prompted by Bob Wilson's
\textquoteleft{}energy saver\textquoteright{} which was a
\textquoteleft{}missing power\textquoteright{} machine---but it later
proved very useful in countering the Member States when they
complained they were in financial straights. In fact it was only when
the final prices came in for the first set of magnets that we knew we
could move directly to exercise an option to order the rest.

\begin{figure}[!ht]
\centering\includegraphics[width=281pt]{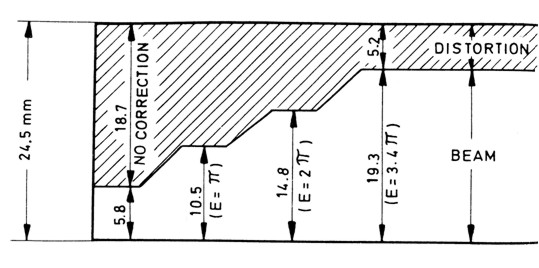}
\caption[]{Correcting orbit distortion liberates aperture}
\label{fig14}
\end{figure}

In the days before computer controls, synchrotrons were designed with
magnet gaps between the poles large enough to accommodate not only the
beam but a generous safety margin to accommodate all the orbit
distortions due to the tiny errors in magnet construction and
alignment with 98\% probability. We invented a strategy based on how
orbit correction had been applied to the PS to liberate aperture by
correcting orbits. By the time the SPS was discussed, the PS had
successfully corrected a large fraction of this orbit distortion,
liberating more aperture for the beam. Why not therefore rely on using
the same kind of correction to reduce the SPS aperture (see
\Fref{fig14})?  I'm not sure if it was my idea but it was one that I
championed.  Perhaps I did not realise it at the time but this was in
danger of pulling the design in the direction of making it less likely
to work first time\textemdash{}one of John's major concerns. However,
it brought about considerable cost savings.

Magnet design is a subject dear to the heart of all
accelerator builders and each (including John) had their idea of how
best to do this. Earlier I explained how Bob Wilson had replaced 1.3 T
combined-function magnets with 1.8 T pure dipoles. But combined-function 
were the magnets John and Kjell knew and loved from the PS and
ISR and had spent many years perfecting. Moreover, some were still
sceptical of Bob Wilson (who the unkind said ran a ranch of cowboys in
the States).

We spent many meetings (then and later when the machine was approved)
discussing the virtues and vices of the new magnet designs.  Many in
the Machine Committee remembered their experiences with similar and
dissimilar magnets that they themselves had built or seen built. It
was perhaps John's biggest challenge to resolve this issue and in the
end it was settled by designing the best lattice for 300~GeV using
combined- and then separated-function principles and looking at the
cost implications using a computer program supplied by the laboratory
that was one of our most vehement critics\textemdash{}Karlsruhe.  John
rightly insisted that everything had to be included in the program. If
the field in a magnet was lowered, the ring became longer and more RF
would have to be added to accelerate in the time defined by the
parameter list. The tunnel would be longer but stored energy which had
to be shipped in and out from the electricity grid would be
reduced\textemdash{}and there were many more such considerations. The
energy dissipated would also change, causing more or less cooling
capacity to be installed. When all this was costed and optimized we
clearly saw that a separated-function ring would cost no more, but
would be more compact. Little did we know at the time that this
matched John's master plan to fit the machine back on the molasse
plateau at CERN, and had the added advantage, vis-\`{a}-vis his
critics, that Bob Wilson's innovations had not been ignored but
exploited.

When all this was over and the Design Report for a 400~GeV machine
written~\cite{ref8}, it turned out that the Machine Committee had done
its job well. The cost savings were important because of the
criticisms of many of the Member States concerning the generous and
expensive safety margins that the Praetorian Guard of old PS designers
had sustained in the \textquoteleft{}300~GeV
Proposal\textquoteright{}. Previous visitors from German and UK
laboratories being shown around the ISR had marvelled openly at the
vast space around the machine\textemdash{}the air
conditioning\textemdash{}gold-plated connections (so it was said) and
the absence of any attempt to learn from earlier experience. The new
design at least seemed to have answered their technical objections.

\subsection{Bringing the 300 back to CERN\textemdash{}`Project B'}

Member States had still to choose somewhere to put it and
Member States were determined to build the next machine anywhere else,
but not at CERN! This was in part fed by the feeling that many
physicists had not succeeded in getting their experiments approved at
CERN while other \textquoteleft{}residents\textquoteright{} had
been preferred. 

You perhaps saw in John Adams's \emph{a posteriori}
analysis of the situation, how, in spite of these objections, it was
his aim to bring the machine back to CERN. Studies of the separated-function 
lattice showed that this might now be possible (at least for
300~GeV). The deciding argument was to be that if the machine were built at CERN it would not
be necessary to set up a whole new laboratory and build a new linac and
an injector synchrotron. The 25~GeV PS was ready and waiting. This idea
came to be known as \textquoteleft{}Project B\textquoteright{}.

For several months in early 1970, Project B had to be kept secret
while he politically manoeuvred the Member States to accept the
idea. They must be attracted by the cost saving.  Every week he set
off each day to a new capital, appropriately dressed and coiffed to
impress the local audience---with the aim of gradually coaxing them
into this new way of thinking.

At first only John and Pepi Dokheer (his secretary) knew
about Project B. However, to check his ideas he had to enlist my help
to calculate the lattice. Unfortunately for both me and Project B, I
had just broken my leg skiing and lay for six weeks with a weight
strapped to my foot in the Cantonal Hospital in Geneva. Computing was
out of the question. I was surprised one afternoon to have a
distinguished visitor at my bedside when John arrived complete with
secretary and chauffeur. He politely enquired about when I might return
to my computer terminal and said that he would have something very
important for me to do when I did.

Sure enough, once I was able to do the calculations, Project B seemed
eminently possible on the CERN site but he still wondered if the
molasse (sandstone) was extensive enough to contain the whole
tunnelling operation and then one day he said, \textquotedblleft{}I
suppose I have to let Jean Gervaise into the secret so that we may
look at the borings in his filing cabinet.\textquotedblright{} (Jean
Gervaise was in charge of surveying the site.)

It was, of course, exciting to work on such a secret project, but
fending off helpful enquires was not easy. Giuseppe Coconni asked me
one day (and I think it was his own idea), \textquotedblleft{}Has
anyone thought of putting the 300 at CERN?\textquotedblright{} I had
to pretend that no-one had considered it, but one might have a look.

Finally, it was time to spill the beans to the Scientific Policy
Committee and then to Council. John cleverly asked Bernard Gregory
(then Director-General) to make the first presentation while he, John,
was in the US, safe from the storm he expected to break, and ready to
return and dampen the flames. As expected, there was quite a lot of
resistance from the physics community who had been hoping for 400~GeV
and, with good luck, closer to their home. There were also many who
probably felt somewhat cheated to hear what had been going on without
their knowledge, and it is debatable whether the secrecy was not
counterproductive.

All this came to a head a few weeks later when the European Committee
for Future Accelerators were asked to approve the new `Project
B'. They met on a Saturday, spending the morning complaining that the
energy was too low, and everything was going rather badly by the time
they adjourned for lunch.  After lunch John took me on one side.  He
had skipped the lunch, returning to his office to ponder over the
cardboard model on his filing cabinet, which showed the contours of
the rock beneath CERN. He said, ``I think we can just find room for an
1100~m radius ring\textemdash{}will this be big enough for
400~GeV?\textquotedblright{} I confirmed that it would, and he offered
it to the afternoon session (with the proviso, to satisfy his
principle of caution, that there were still some crucial borings to be
done which might yet bring a nasty surprise). It was enough to turn
the tide in his favour and save the day for Project B.

There were still many Member States to be convinced to join. This took
until the Council meeting of December 1970 and even that had to be
adjourned and reconvened on 19th February 1971 before the last couple
of Member States could be persuaded. That afternoon, after a
particularly good Council lunch, John lost no time in returning to his
office to start the business of recruiting the new staff. There were
600 farmers who owned the land on which the new ring was to be
built. His first appointment that afternoon was with Andr\'{e} Klein,
a high official from the Prefecture of the region whom he persuaded to
join the team to deal with any dissent from the landowners.

\subsection{Highlights of SPS construction}

The offices of the new Laboratory II were in a barrack as far from the
centre of CERN as possible, and later were moved over into France near
Pr\'{e}vessin. It was clear he wanted to put his imprint on a new
style. Again he had only one weekly technical meeting, like the
Parameter Meetings he had chaired for the PS. He chaired this Design
Committee every Monday morning until the machine was finished. The
team he assembled over the few weeks following SPS approval was a
healthy mixture of those who had helped him with the PS and who had
gone on to build the ISR, and new blood from other Member State
laboratories.

\begin{figure}[!ht]
\centering\includegraphics[width=341pt]{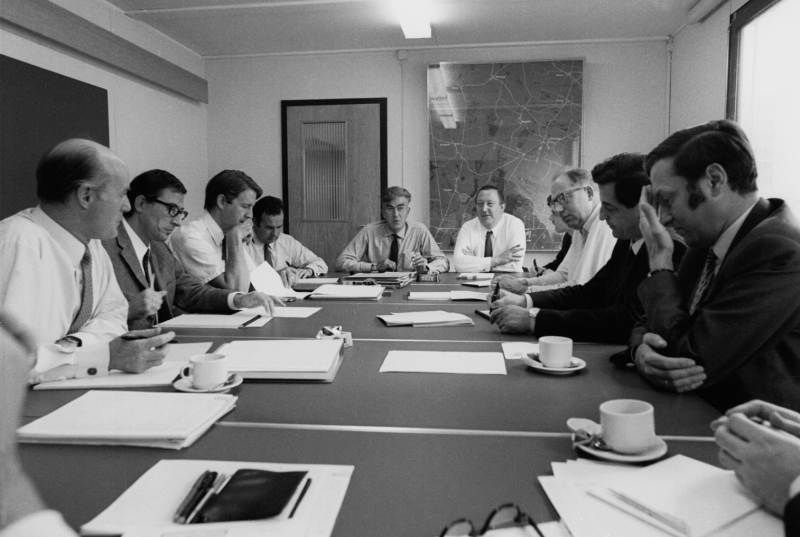}
\caption[]{The 300~GeV Design Committee}
\label{fig15}
\end{figure}

\Fref[b]{fig15} is a photograph taken on the occasion of the first
meeting of the 300~GeV Design Committee from my viewpoint opposite
John. On his left is his second-in-command Hans-Otto Wuester, a
charismatic but explosive German from DESY Hamburg who had, as he
reminded anyone who was slow to respond to his encouragement,
\textquotedblleft{}one shoe that is sharpened to be used where it
hurts.\textquotedblright{} He was the foil to John's gentlemanly
manner and was often sent over by John to Laboratoy I to
\textquotedblleft{}sort them out\textquotedblright. Usually the threat
was enough!

We also see, going round clockwise from the left, Hans Horisberger
(engineering), Clemens Zettler (radio frequency), Roy Billinge
(magnets), Norman Blackburne (personnel), Bas De Raad (extraction),
Klaus Goebel (health and safety), and Simon van der Meer (power
supplies). Others who came later included Boris Milman (finance and
planning), Giorgio Brianti (experimental areas), Michael
Crowley-Milling (control system), and Robert L\'{e}vy-Mandel (civil
engineering).  Wuester, Billinge, Milman, L\'{e}vy-Mandel and
Crowley-Milling came from outside.  Others: Zettler, Blackburne, and
Goebel, were second-in-command to CERN group leaders who presumably
chose to stay where they were.

\begin{figure}[!ht]
\centering\includegraphics[width=270pt]{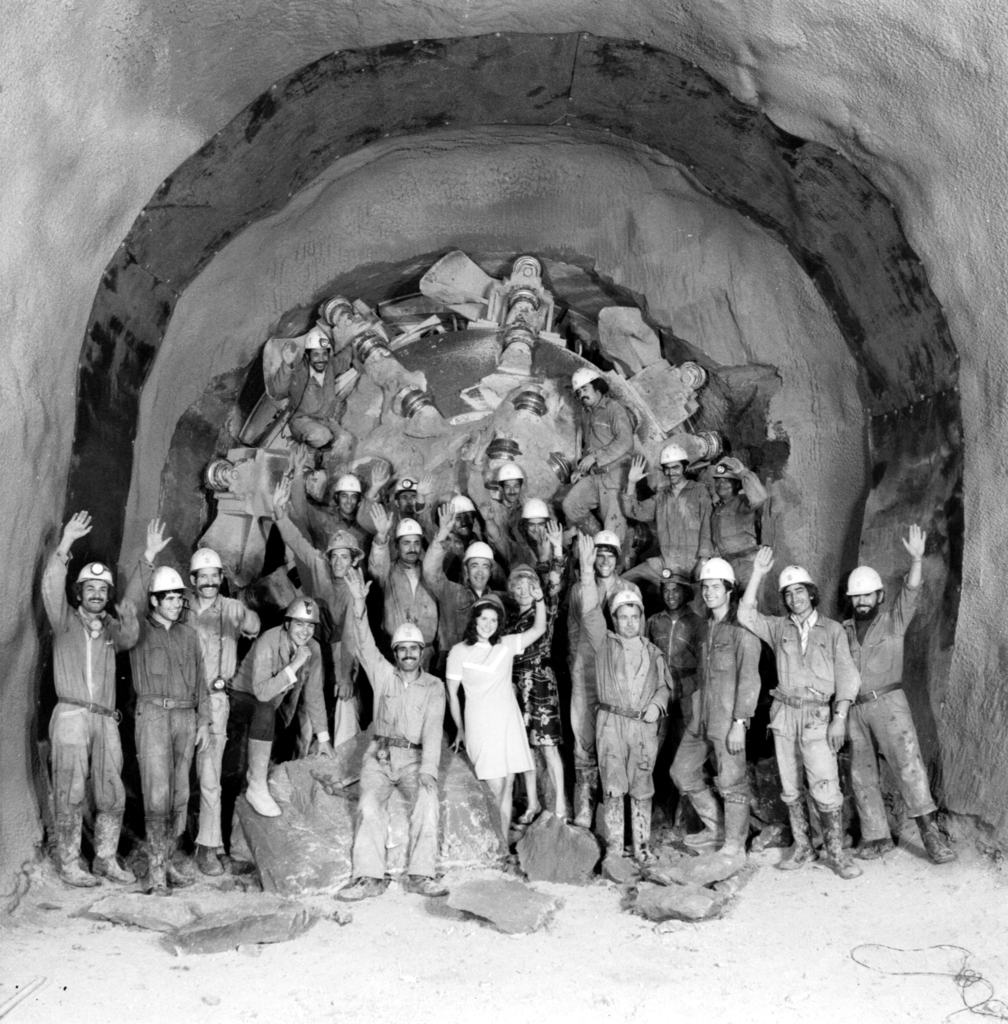}
\caption[]{The SPS ring tunnel is completed}
\label{fig16}
\end{figure}

John was particularly interested in keeping an eye on civil
engineering. On Saturdays he would tour the site with Robert
L\'{e}vy-Mandel, noting where work might be falling behind schedule. By the
time Monday came around again Robert would usually be able to report
that he had talked with the contractors and found a solution. Placing
large contracts for the magnets was another major concern. If the
second half of the magnets for SPS were to be ordered, the contract for
the first half would have to come in at a low price and options to build
the second half would have to be written into the agreement.

\subsection{Magnet problems}

There were from time to time, as in any project, unforeseen
technical setbacks. One such was the discovery that 100 of the 700
bending magnets already installed in the tunnel had developed short
circuits to ground. This was deeply shocking to all concerned and it
looked as if the SPS was no better than the Fermilab main ring where
magnets failed at the rate of one a day in the early tests. Had we been
wise to emulate the methods of Fermilab? we wondered. 

The whole team of group leaders was summoned to meet every
day for a week to investigate the cause and plan a remedy. It was in
the spirit of putting their heads together. There might well have been
shouting or admonishment, but with John Adams in the chair there was
instead, as there should be, just logic and science. 

Rather soon Billinge and Bob Sheldon, who was a chemist, and whose
first instinct was to lick a finger and to taste the tag ends of the
coil conductors, established that they had been prepared for brazing
by cleaning with phosphoric acid by an overzealous welder. The acid,
it was discovered, could fill up the hollow glass fibres which loaded
the insulation and provide a conducting path for short circuits.
Fortunately there was time, without delaying the start-up, to take out
the infected magnets, rebuild the coils, and wrap them in Kapton to
prevent any other shorts.

Delays to several large projects (not least, the magnet insulation of
the Fermilab Main Ring, the niobium welds on the vacuum chamber of the
Large Electron--Positron machine (LEP), the busbar connections of the
Tevatron, and recently the interconnects in the Large Hadron Collider
(LHC)) have regularly been caused by the unpredictable consequences of
engineering solutions.  It was fortunate that no delay resulted for
the SPS\textemdash{}perhaps thanks to John Adams's rigourous analysis
of possible difficulties and their solutions\textemdash{}but more
realistically because of the thorough pre-start-up tests he insisted
upon after installation in the tunnel.

\subsection{Commissioning}

The SPS was finished five years after the team first
assembled in Pr\'{e}vessin. Such was the thoroughness of the preparation
that John had expected from his team that each stage in the
commissioning programme worked like clockwork. Once again John left those
in charge to do their stuff, but I do remember one moment before the
beam was injected when he asked everyone ``Are you sure you have not
forgotten something?''. 

\begin{figure}[!ht]
\centering\includegraphics[width=200pt]{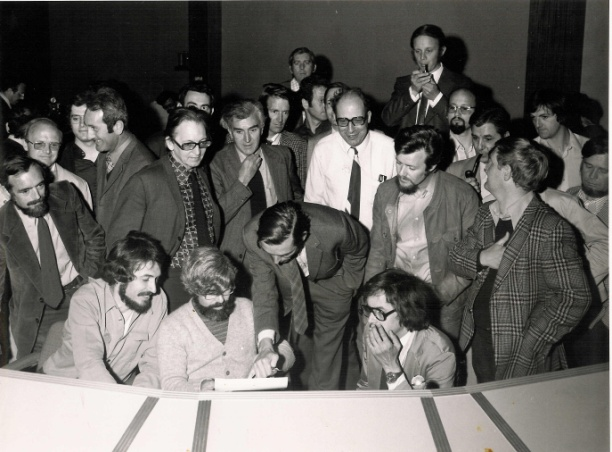}
\caption[]{SPS control room---first beams accelerated}
\label{fig17}
\end{figure}

The contrast with the commissioning of Fermilab, which I had
lived through a couple of years earlier, was clear. Everyone in the SPS
control room had done their professional job and knew enough about
accelerator physics to diagnose any little misbehaviour of the beam. To
be fair, it also helped to be able to learn from Fermilab experience. There
was one little hiccup when we tried to accelerate for the first time
and the beam just disappeared. Within the same day we tracked down a
fault in the numerical program of the power supplies for the
focusing system and went on to accelerate.

\begin{figure}[!ht]
\centering\includegraphics[width=199pt]{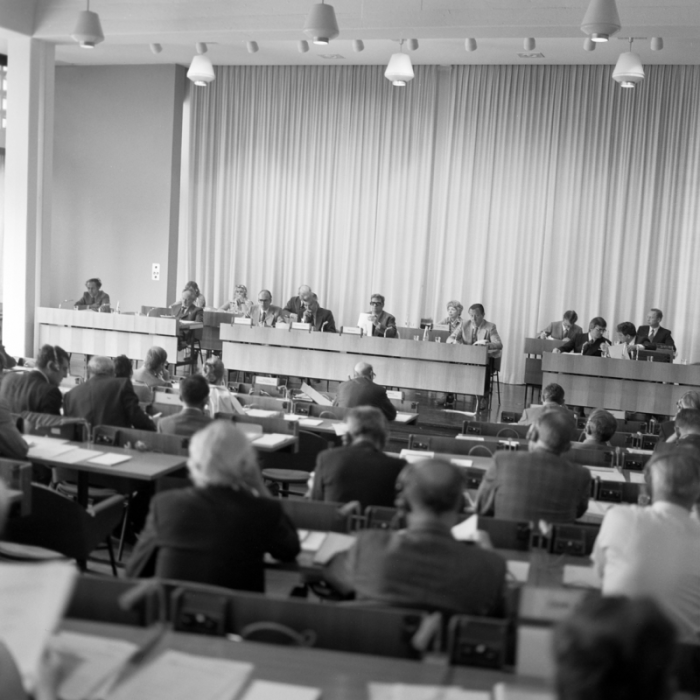}
\caption[]{The CERN Council is asked to approve 400~GeV}
\label{fig18}
\end{figure}

The 200~GeV acceleration came easily and the first pulse was
synchronized to be announced to the Council at the end of their morning
session. At some time in the past, the Council had insisted they be
asked permission before moving from 200~GeV to 400~GeV. It had been
something to do with ordering the missing magnets. After reporting
acceleration to 200~GeV, John wryly asked their authorization to
accelerate to 400~GeV and by the tea break in the afternoon he announced
the first 400~GeV pulse---on time and of course---on budget.

\begin{figure}[!ht]
\centering\includegraphics[width=204pt]{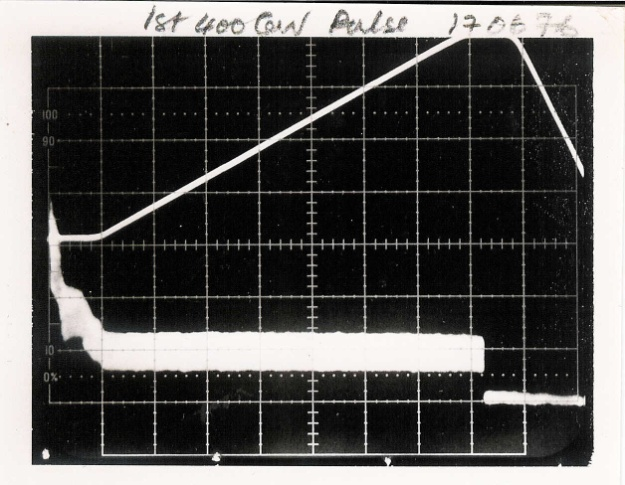}
\caption[]{The first 400~GeV pulse}
\label{fig19}
\end{figure}

\subsection{He becomes Director-General a second time}

During the construction period he had been Director-General of
Laboratory II, which included the SPS and the Pr\'{e}vessin site in
France, while Willi Jenschke had looked after the main Laboratory I
site at Meyrin including the ISR, PS, and their experiments. The time
came, just before the SPS was finished, to merge the two laboratories
together under a single Director-General. I remember meeting him then
in the corridor (during the Council meeting where this was to be
decided). He confided in me disconsolately that,
\textquotedblleft{}they were taking a long time over it---for some
reason I do not understand, they think they need to have two
DG's.\textquotedblright{} It turned out that, while they recognized he
was the man to look after the accelerators, they wanted an eminent
physicist rather than an engineer to manage the research
programme. And so it was that they decided that John would be one
Director-General who would concentrate on the accelerators, while
L\'{e}on van Hove, a second Director-General, looked after the physics
programme. It is greatly to John's credit that he was able to accept
this arrangement and together they made it work. During his final term
as Director-General he visited China. It was 1977 and before the iron
grip of the Gang of Four began to slacken. China was keen to build a
large proton ring near the Great Wall as a statement of China's
progress towards western prosperity. John met Deng Xiaoping.
\textquotedblleft{}Very smart,\textquotedblright{} said John,
\textquotedblleft{}perhaps I made a mistake to tell him the big proton
machine would be no use to them and what they really needed was a
synchrotron light source.\textquotedblright{} And of course that is
what they did.

\subsection{R. R. Wilson}

Throughout the construction of the SPS, Robert Rathbun Wilson was
John's US counterpart whom he rarely mentioned. Bob Wilson had set
about constructing the Fermilab main ring with very similar design
aims to the SPS.  He was fortunate to be able to start about five
years before SPS approval and had finished it (though there were still
some things to tidy up) about five years before the SPS.  His style
could not have been more different from that of John Adams. I had the
privilege of working closely with both these men for, in the middle of
SPS construction, I was dispatched by John Adams to help sort out some
of the difficulties that Bob Wilson was having in commissioning his
400~GeV Main Ring. The fact that it needed someone from another
laboratory to help in this way is perhaps a comment on the risks that
Bob Wilson was prepared to take to save time and money in
construction.  This was something that John Adams, in his desire to be
careful and not prejudice the reliable operation for the machine, was
at pains to avoid.

However, it must be said that Bob Wilson inspired younger
members of his team (and ours) with his bold initiatives. Many of the
ideas which simplified the design of the SPS and assured its success
had been copied from innovations he pioneered at Fermilab.  We have seen
that John Adams had embraced these ideas with enthusiasm, provided they
did not put the outcome at risk.

Once, while visiting Fermilab, I can remember being asked by a
resident historian to compare these two great men. My answer was:

John Adams had artistic talent but had never had the time to follow
his talent to its conclusion\textemdash{}Bob Wilson on the other hand
had managed to achieve an international reputation as a sculptor and
architect.

John Adams persuaded through reason and was always a gentleman
\textemdash{}Bob Wilson challenged his team with his own inspiration
and rode roughshod over their objections.

John Adams was careful\textemdash{}Bob Wilson deliberately took
risks (but was prepared to fix them afterwards).

John Adams was ideal for Europe whose politicians are used to allowing
themselves to be persuaded by the reason and common sense of their own
scientific advisors. Bob Wilson's passionate rhetoric often rivalled
the Fathers of his Nation and was finely tuned to the ear of a
Washington politician or media magnate.

Both would have been a disaster had they exchanged the old world for
the new, or vice versa. Perhaps John Adams would have found it even
easier to establish his technical dominance in the USA and without
formal qualifications, but the US has little time for the staff
management methods of Cockcroft or the cerebral exercises of Socrates. 
Bob Wilson would probably have been judged rash by European politicians
and scientists, but his artistic gifts would have found more nourishment
in the richer soil of Paris, Florence, or Rome than in Illinois.

\emph{Both felt their career should have gone on longer, and I agree!}

\section{Since he left us \ldots{}}

In preparing this talk I was asked to answer the question
 \textquotedblleft{}How would CERN be different if John Adams was
 still alive today?\textquotedblright{} I will attempt to answer this,
 but emphasize that this is merely a personal view. I have not been as
 closely involved in CERN's recent projects as I was when John Adams
 was alive and I expect that those who were may disagree with my
 conclusions. I still think that the points I raise are worthy of
 debate and should be taken on board by leaders of future projects.

The first accelerator project to follow in the wake of his years as
Director-General was the Antiproton Accumulator (AA) using the SPS for
colliding antiprotons with protons. This involved also the
construction of two large detectors, UA1 and UA2. John Adams was still
with us when these projects and LEP were started, and accelerator
engineers and physicists who had been schooled in his way of doing
things were largely responsible for their execution. These projects
still bore his footprint.

The development of a more intense antiproton source to follow AA was
perhaps something he might have restrained, given that the Tevatron
with twice the centre-of-mass energy of the SP-PbarS was about to put
antiproton physics with the SPS out of business. However, the cost of
the new Antiproton Accumulator turned out later to be a small price to
pay for the improved supply of antiprotons for the low-energy LEAR
programme. About this time there was an upgrade to UA1 which did not
materialize. John Adams might have seen this coming but would probably
not have been able to restrain it even if he had wanted, since it was
outside the field of accelerators.

I am tempted to think that the teething troubles due to the use of
magnetic material (niobium) in the finishing of the LEP vacuum chamber
might have been prevented by a Design Committee with John to guide
them\textemdash{}or maybe it was just bad luck. Anyway, the delay it
caused was minimal and LEP proved a great success in spite of it.

LEP had expensive delays due to fountains of water springing from the
walls and floor during the tunnelling. With hindsight it should never
have encroached upon the Jura limestone. John would certainly have
been aware of the dangers of leaving the molasse and tunnelling into
water-bearing rock but it would have required all his skills to
persuade the physics community to sanction a smaller and less
energetic LEP.

The next phase that John Adams might have had an influence upon was
the race for approval between the Large Hadron Collider (LHC) and the
Superconducting Supercollider (SSC). Once the US and Texas had decided
they could not foot the bill for the SSC, he would have been in his
element trying to arrive at a machine which the world might
afford. Had this come to pass and had he gone on to have a leading
role in a Super Collider's construction, he would have kept a tight
control on the tenders for major hardware components\textemdash{}a
scrutiny which was very much needed at the SSC. Both the SSC and LHC
used superconducting magnets, and it would have been interesting to
see if John Adams could have found a way to curb the fears of
industrial firms whose tenders for superconducting magnets mainly
reflected their caution in bidding for an unfamiliar technology.

Approval for the LHC took a long time, but then so did approval for
the SPS.  After the demise of the SSC, Chris Llewellyn-Smith and
Giorgio Brianti finally took the Council by the horns, and got them to
agree to the LHC. Their approach used many of the techniques that
Adams had deployed in 1971 to secure the SPS for CERN.

As for the future linear collider, I like to think that John would
have seen the virtue of a common cause which spanned the various
laboratories involved earlier, and used collaboration to push CLIC
more rapidly towards becoming a project rather than a research and
development exercise.

Of course, the big question at the moment of writing is---Would John
and his Design Committee have seen the troubles with the LHC
interconnects which caused so much sorrow in the last twelve months?
This is in many ways reminiscent of the SPS history of magnet
insulation problems. The only difference perhaps is that the SPS
problem became apparent during routine electrical tests rather than in
the glare of the spotlight of the world press. Nevertheless, there is
a strong probability that John (always on the lookout for engineering
weaknesses) would not have let it creep under the radar of his Design
Committee.

\section{Conclusion}

It was in 1981 that Sir John Adams received his knighthood from the
Queen, but he modestly never asked to be called Sir John by his
colleagues. Once his term of office was over, he moved back to his old
office on the Pr\'{e}vessin site and began to make himself available
as an advisor to a number of European and other international
bodies. He would really like to have built LEP but as he said,
\textquotedblleft{}Schopper was keen to do it.\textquotedblright{} His
brilliant career was at an end, and in the last few years he missed
the bustle of building accelerators and the long queue of those
waiting to see him, but I suppose that comes to all as they approach
retirement, and what a career he had had! And what a legacy he left
behind at CERN! There is so much in his career that those at CERN
would do well to remember every time they start a new accelerator
project. Not all of us can have his gifts but we may aspire to them.